\documentclass[twocolumn]{aastex631}
\usepackage{amssymb,amsmath,graphicx,multirow,float,footmisc,natbib,hyperref,placeins}

\def\gtrsim{\mathrel{\hbox{\rlap{\hbox{\lower4pt\hbox{$\sim$}}}\hbox{$>$}}}}
\def\lessim{\mathrel{\hbox{\rlap{\hbox{\lower4pt\hbox{$\sim$}}}\hbox{$<$}}}}

\begin{document}

\title{Short-Duration Gamma-Ray Burst and Afterglow Rates in the Rubin and Roman Era}

\author[0000-0002-4319-1615]{Tzvetelina A. Dimitrova}
\affiliation{Earth and Space Exploration, Arizona State University, P.O. Box 871404, Tempe, AZ 85287-1404, USA}

\author[0000-0002-9110-6673]{Nathaniel R. Butler}
\affiliation{Earth and Space Exploration, Arizona State University, P.O. Box 871404, Tempe, AZ 85287-1404, USA}

\begin{abstract}
Short-duration gamma-ray burst (sGRB) afterglows that follow BNS-gravitational wave (GW) events are essential for understanding the tension between the observed sGRB rate and BNS merger rate, heightened by the recent conclusion of aLIGO O4 with no new confirmed BNS detections. Using a probabilistic sGRB world model derived from a source BNS merger population, we simulate afterglow emission with \texttt{AfterglowPy} to investigate detection prospects of afterglows in the new era of optical surveys, and probe their multi-messenger implications. The predicted sGRB/BNS association is strongly dependent on sGRB beaming, which may be constrained by orphan afterglows (OA) - that arise from events with no prompt $\gamma$-ray detection. We conclude that the Vera C. Rubin Observatory's Large Synoptic Survey Telescope (LSST) may detect an afterglow sample sufficient in constraining sGRB jetting, with an estimated $5.3^{+1.7}_{-1.2}$ on-axis afterglow and $11^{+5}_{-3}$ OA detections per year; while the deep sensitivity of the Roman Space Telescope appears promising for probing the faint end of afterglow events in targeted follow-up strategies. The detection of afterglows in upcoming LIGO runs is possible but challenging, as we predict fewer than $\approx 1.4$ LSST or Roman discoverable events per year within the projected aLIGO O5 BNS range across all considered jet models and observing scenarios. We update previous sGRB-BNS rate predictions, finding that continued non-detection of a BNS in O5 would require revisiting key assumptions underlying sGRB-BNS models.
\end{abstract}

\section{\textbf{Introduction}} \label{sec:intro}
The discovery of gravitational wave (GW) GW170817A with a coincident electromagnetic (EM) detection of short gamma-ray burst (sGRB) GRB170817A from a binary neutron star (BNS) coalescence marked the dawn of multi-messenger astronomy. However, in the most recent aLIGO observing run (O4), there have been no additional detected BNS events. Here, we investigate the role of sGRBs afterglows (AG), including orphan afterglows (OAs), following BNS-GW events by predicting their rates and expected distributions of observables for the next generation of optical surveys - and also estimating detection rates for the upcoming aLIGO O5 run \citep{LIGO2025ObservingScenarios}. 

OAs are the afterglows (AGs) of GRBs seen without prompt $\gamma$-ray jet emission, which may occur through two different channels. Off-axis orphans are those viewed beyond the main core of the jet, later potentially seen as afterglows when the jet decelerates and widens into the interstellar medium.
Such afterglows are thought to be greatly more numerous than the current $\gamma$-ray band detected GRB population, with $\sim 2/\theta_{jet}^{2}$ off-axis bursts expected for every typical GRB jet opening angle $\theta_{jet}$ \citep{Ghirlanda2015}. 
In contrast, faint OAs correspond to GRBs too faint for satellite detection (off-axis or not) but with afterglows sufficiently bright for optical detection. 

Strong tension exists between the detected rate of sGRBs and the estimated rate of their likely BNS progenitors. In \cite{Dimitrova_2023}, we employed a probabilistic sGRB world model to predict joint Swift sGRB/aLIGO detection rates from BNS mergers, concluding that the predicted number of sGRB/GW associations strongly depends on sGRB beaming and highlighting the importance of understanding the unique off-axis nature of GRB 170817A. Constraining sGRB jetting effects is essential for understanding whether the observed sGRB rate is consistent with the observed BNS rate such that merger events can form the source population for sGRBs.

Here, we study the detection prospects for OAs in the emerging era of deep and wide-field optical sky surveys. 
The Vera C. Rubin Observatory's Large Synoptic Survey Telescope (LSST) is expected to achieve a single-visit sensitivity of $m^\mathrm{AB}_{r} \approx 24.7$ in 30s ($5\sigma$) and exceptionally wide sky-coverage for potential wide-field afterglow detection \citep{Ivezic2019}.
Promising for probing the fainter afterglows, the Nancy Grace Roman Space Telescope \citep{Roman} is set to launch in mid-2027. This optical/NIR mission would be advantageous due to a deep sensitivity of $m^{\rm AB}_{RO62} \approx 27.97$ in 1-hour
($5\sigma$) \citep{RomanWFI}. 

These facilities provide a marked sensitivity increase relative to other existing surveys 
(e.g., the Zwicky Transient Facility; ZTF \citep{ZTF}, with a 30 s, $5\sigma$ limiting magnitude of $m^\mathrm{AB}_{r} = 20.6$).
By simulating afterglows for each sGRB in our world model, we quantify potential detection rates and discuss optimal survey strategies. From the recovered observable distributions (e.g., for redshift $z$, isotropic equivalent energy release $E_{\rm iso}$, and jet angle, Figure \ref{fig:obs_dist}) we also study the extent to which future OA detections can provide constraints on fundamental sGRB parameters. 

\subsection{\textit{Previous Estimates}}
The rate of OAs in optical surveys is poorly constrained. \cite{Lamb2018} estimate an LSST detection rate of $16 - 76\, \rm yr^{-1}$ and a ZTF rate of $2- 8\, \rm yr^{-1}$ from jets of $6$$^{\circ}$, $16$$^{\circ}$, and $26$$^{\circ}$ within $z = 0 - 3$. \cite{Ghirlanda2015} find upper limits of $ < 50$ LSST OAs $\rm yr^{-1}$, and $< 20$ $\rm yr^{-1}$ ZTF OAs. Due to their intrinsic faintness and delayed magnitude peaks - from days in the optical \citep{Ghirlanda2015}) to years post burst in radio \citep{Ghirlanda2014} - OA observation requires deep detector sensitivity and careful timing. Few OA candidates currently exist, none of which are definitively confirmed, as distinguishing them from other slow transients presents further challenge \citep{Nakar2002, Leung2023}. In \cite{Perkins2024}, \texttt{AfterglowPy} \footnote{\label{foot:afterglowpy}\url{https://afterglowpy.readthedocs.io/en/dev/modules/afterglowpy.html\#id1}} \citep{Ryan_2020} is similarly used to investigate BNS detection prospects through optical channels. Assuming that all mergers launch sGRB jets and that observations are obtained at peak afterglow brightness, they find that the inclusion of nearly on-axis afterglows increases the LSST $g$-band BNS merger discovery rate from $29\, \rm yr^{-1}$to $\approx 91\, \rm yr^{-1}$.

Off-axis sGRB events additionally face detection difficulty, with existing jet constraints remaining broad and often only providing lower limits of sGRB jetting (typically ranging from $\theta_{\rm jet} \ge 3^{\circ}$ to $\theta_{\rm jet} \ge 20^{\circ}$ \citep{Berger2014}) rather than precise measurements of opening angle, which depends on the rare observation of a jet break. \cite{Ghirlanda2016} derive an estimated $3^{\circ} - 6^{\circ}$ average $\theta_{\rm jet}$, while \cite{Dimitrova_2023} models suggest a broader $\theta_{\rm jet} \gtrsim 30^{\circ}$ distribution for nearby events as compared to cosmological sGRBs. 

GRB170817A further emphasizes the importance of tightening such constraints. With a narrow off-axis jet (likely a viewing angle of $\theta_{\rm view} \sim 5 \theta_{\rm jet}$, \cite{NakarPiran2021}) despite a low $E_{\rm iso} = 5.35 \times 10^{46}$ erg suggesting a nearly spherical explosion \citep{Dimitrova_2023}, it's unique natures indicates a key clue to understanding sGRB-GW association. 

\begin{figure*}[ht!]
\centering
\includegraphics[width=1\textwidth]{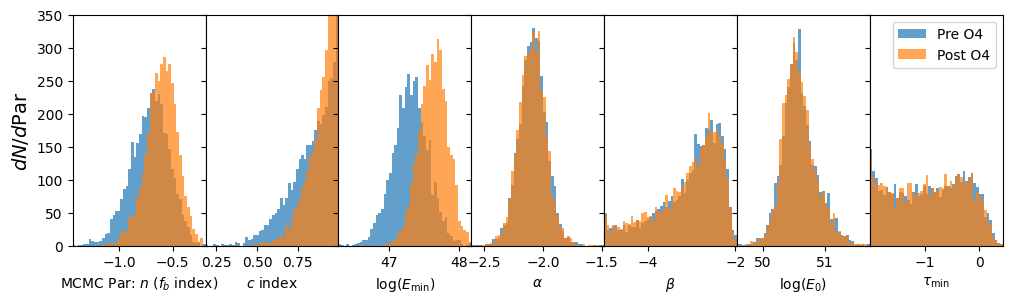}
\caption{sGRB population model MCMC parameter distributions before the non-detection result of O4 (blue), and post-O4 analysis (orange).}
\label{fig:mcmc_params}
\end{figure*}

\section{\textbf{An Updated sGRB World Model}} \label{sec:data}
We begin by re-fitting the sGRB World model from \citet{Dimitrova_2023}, including
updated estimates for the durations and distances probed by the prior aLIGO runs.
Based on the various start and stop dates throughout each run, we take a total
duration for the aLIGO runs through O3 of $t_{\rm O2+O3} = 1.72\, \rm yr$ with a corresponding average distance of $D_{\rm O2+O3} = 110$ Mpc. For the O4 run, we assume $t_{\rm O4} = 1.89\, \rm yr$ and $D_{\rm O4} = 160$ Mpc \citep{LVK}. As in \citet{Dimitrova_2023}, we assume 2 BNS events have been detected, and we require the world model to also fit the sGRBs detected by {\it Swift} between 2004 December 17 and 2020 August 29.  

The aLIGO O4 run increases the time-volume product relative to the O2 and O3 runs by a factor of approximately 4.4, which drives the predicted BNS rate substantially lower. As shown in Figure \ref{fig:mcmc_params}, in order to reproduce the observed number of sGRBs, the model including the O4 constraints must have a flatter beaming distribution, with $P(f_b) \propto f_b^n$ (i.e., more sGRBs pointed toward us) where the beaming fraction is $f_b = 1 - \cos(\theta_{\rm jet})$; and also a larger minimum energy cutoff, $E_{\rm min}$, in the luminosity function (i.e., a larger fraction of bright, detectable sGRBs). 

The luminosity function operates on $E=E_{\rm iso} f_b^c$, and an increase in the parameter $c$, which is highly-correlated with $n$, towards unity indicates that the luminosity function operates primarily on the beaming-corrected energy release, $E_{\rm iso} f_b$, as opposed to $E_{\rm iso}$. The parameters corresponding to these changes are those in the first three panels of Figure \ref{fig:mcmc_params}, and the remaining parameters (i.e., the luminosity function parameters $\alpha$, $\beta$, and $E_0$, and the rate density parameter $\tau_{\rm min}$) are effectively unchanged. The best-fit parameters and confidence intervals are given in Table \ref{tab:mcmc}. 

\begin{figure}[!hbt]
\centering
\includegraphics[width=1\columnwidth]{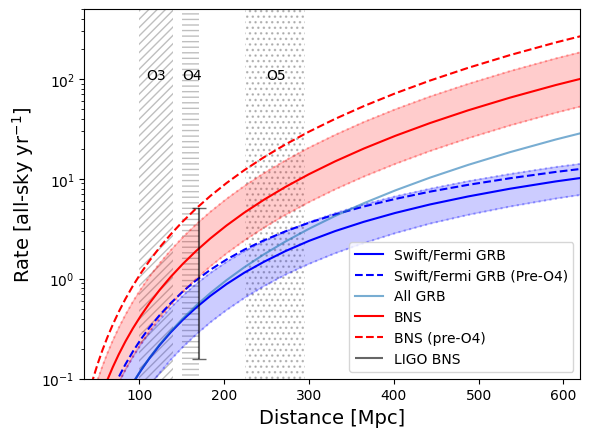}
\caption{Predicted cumulative all-sky per year rates post-O4 correction for Swift/Fermi sGRBs (blue), the full sGRB population (light blue), and BNS mergers (red) as a function of distance; dashed curves show corresponding pre-O4 analysis results. Gray shaded regions mark representative BNS ranges of aLIGO O3, O4, and O5 \citep{LVK}. The aLIGO BNS merger rate estimate \citep{Abbott_2025_GWTC4_Population} is scaled to an upper O4 BNS limit (black). Shaded bands denote $90\%$ confidence intervals for the post-O4 sGRB and BNS rates.}
\label{fig:grb_rate}
\end{figure}

Figure \ref{fig:grb_rate} displays our predicted BNS and joint BNS$+$sGRB cumulative rates as a function of distance, with the distances corresponding to various past and upcoming aLIGO runs marked in gray. Our BNS estimates including O4 are higher on average than those from aLIGO alone (``aLIGO BNS'' in Figure \ref{fig:grb_rate}) \citep{Abbott_2025_GWTC4_Population}, because the demand that the BNS population source the sGRB population props up the BNS rate. Table \ref{tab:grb_rate} displays our predicted cumulative sGRB and BNS rates in GW-BNS relevant distances. In Section \ref{sec:aLIGO_implication}, we discuss the likelihood of this rate yielding no detections in the aLIGO O4 run and the (extremely unlikely) possibility that it should again yield no detections in the upcoming aLIGO O5 run.

\begin{table}[h!]
\renewcommand{\arraystretch}{1.2}
\centering
\caption{\textbf{MCMC Parameters ($\mathbf{90\%}$ Conf. Errors)}}
\label{tab:mcmc}
\begin{tabular}{ll|ll}
\hline
\hline
\textbf{Parameter}   & \textbf{Value}  & Parameter & Value  \\ \hline
$f_b$ index, $n$   & $-0.6^{+0.2}_{-0.3}$ & $\alpha$             & $-2.1 \pm 0.2$\\
$c$    & $> 0.8$ & $\beta$             & $-2.9_{-1.7}^{+0.7}$ \\
$\log\,(E_{\rm min})$  & $47.6^{+0.3}_{-0.4}$ & $\log\,(E_0)$   & $50.6^{+0.5}_{-0.4}$ \\
& & $\tau_{\rm min}$                  &  $< 900\, \rm Myr$ \\
\hline
\hline
\end{tabular}
\end{table}

\subsection{Afterglow Simulations}
To determine afterglow rates, we need to determine an effective beaming fraction $f_{b, \rm eff}$ -- the fraction of sky over which an afterglow can be detected as a function of observation time $t_{\rm min}$ and approximate $r$-band survey limiting magnitude $m_r^{\rm lim}$ -- for each sGRB in our world model. We first bin the sGRB world model counts in terms of $z$, bolometric fluence $S$, and beaming fraction. We then simulate afterglows for each bin, over a range of circumburst densities $n_{0}$, for two jet types (top-hat and Gaussian), and over a range of viewing angles extending beyond the edge of the top-hat jet. For each afterglow, by integrating over viewable angle, we calculate an effective beaming fraction:
\begin{equation}
\begin{aligned}
f_{b,\,\mathrm{eff}} =
\int &\;
\Theta\!\{m_r^{\rm lim} - m_r(\theta_{\mathrm{view}}, t_{\min}) \}  \sin(\theta_{\mathrm{view}})\, d\theta_{\mathrm{view}},
\label{eq:fbmax}
\end{aligned}
\end{equation}
which replaces $f_b$ in order to map sGRB rate to afterglow rate. In Equation \ref{eq:fbmax}, $\Theta\{\}$ is the Heaviside function, and $m_r$ is afterglow peak $r$-band magnitude. When applying $m_r^{\rm lim}$, we also average over Galactic extinction, with $A_r$ assumed to be exponentially-distributed with mean $A_r = (1+z)/4$. As in \citet{Dimitrova_2023}, in order to obtain a self-consistent mapping between $\theta_{\rm jet}$ and $f_b$ that allows us to utilize the rates determined assuming top-hat jets, we assume that the core of the Gaussian jet $\theta_{\circ} = \tan(\theta_{\rm jet})$, where $\theta_{\rm jet}$ is the half-width of the top-hat jet.

We generate the $r$-band afterglow light curves using \texttt{AfterglowPy}. The software calculates the observed flux density by integrating the specific intensity of synchrotron emission over solid angle at observer-frame time $t$, including power-law scaling relations for the evolution in $t$ and frequency.
The calculation takes as input parameters describing the sGRB beaming, energetics, and environment. The beaming model is specified by jet type (e.g., top-hat, or Gaussian). The flux normalization is set by $E_{\rm iso}$ and $z$ from the world model. We assume default values for the microphysical parameters: electron energy distribution index $p$ (assumed 2.2), fraction of electrons accelerated $\chi_{\rm N}$ (assumed 1), and the fraction of total shock wave energy allocated to electron kinetic energy and magnetic field ($\epsilon_{\rm e}$ and $\epsilon_{\rm b}$, assumed to be 0.1 and 0.01, respectively). 

The microphysical parameters have been found to vary significantly for sGRB afterglows ($\epsilon_{\rm b}$ in particular by 2--3 orders of magnitude), with best-fit density values varying by 5--6 orders of magnitude \citep[see, e.g.,][]{Fong2015}.  Although it is important to allow the microphysical parameters and density to vary for broadband modeling or when fitting individual afterglows, the microphysical parameters and density are highly-degenerate in setting the normalization at $r$-band alone (for a given $E_{\rm iso}$ and jet geometry).  In order to generate a plausible sample of afterglows, and seeking to avoid systematics that could arise from mistreatment of any intrinsic correlations between these parameters, we choose to adopt fixed microphysical parameters and allow only the density to vary widely.

\begin{figure}[hbt!]
\centering
\includegraphics[width=1\columnwidth]{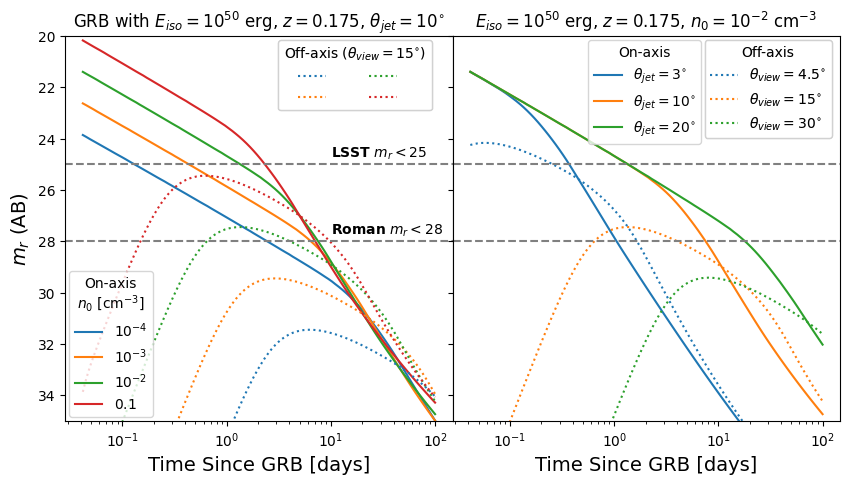}
\caption{Simulated sGRB $r$-band afterglow light curves in a top-hat jet model, assuming fixed values representative of a typical observed sGRB, for on-axis (solid) and off-axis (dotted) viewing angles. LSST and Roman $r$-band detection thresholds are overlaid (gray, dashed). \textit{ Left:} $E_{\rm iso}$, $z$, and $\theta_{\rm jet}$ are fixed to show light curve dependence on circumburst density. \textit{Right:} $E_{\rm iso}$, $z$, and $n_0$ are fixed to show light curve dependence on jet geometry.}
\label{fig:afterglow_param}
\end{figure}

In Figure \ref{fig:afterglow_param} we display light curves of a representative observed sGRB (adopting fixed values typical of the observed population, see Figure \ref{fig:obs_dist}; \citealt{Fong2015}) for on- and off-axis viewing cases, and different input values of circumburst density. Density is the primary parameter affecting the afterglow brightness for a given $E_{\rm iso}$ and $z$ (see Section \ref{sec:param_dep}). Guided by \citet{Connor2020}, we simulate afterglows over a range of densities between $10^{-4}$ cm$^{-3}$ and 10 cm$^{-3}$. When averaging over densities, we assume a split log-normal distribution with a mean log-density of -1, extending gradually to smaller values assuming a standard deviation of 3/2 and rapidly cutting off at larger values assuming a standard deviation of 3/4.  Also following \citet{Connor2020}, we assume an efficiency $\epsilon_{\gamma}=0.15$ for the conversion of bulk kinetic energy to gamma-rays for the sGRB.

\begin{figure}[!t]
\centering
\includegraphics[width=1\columnwidth]{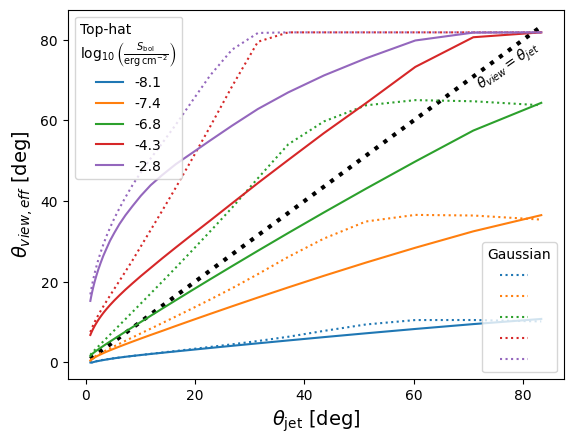}
\caption{The relationship between sGRB jet opening angle and afterglow effective detectable viewing angle $\theta_{\rm view, eff}$, where $f_{b, {\rm eff}} = 1 - \cos{(\theta_{\rm view, eff})}$ (i.e, the maximum viewing angle from which the afterglow remains detectable) at several $S_{\rm bol}$, for both a top-hat jet model (solid) and Gaussian jet model (dotted).}
\label{fig:tj_tview}
\end{figure}

To visualize how Equation \ref{eq:fbmax} establishes afterglow rates from input sGRB rates, it is useful to plot $f_{b,{\rm eff}}$ versus the sGRB jet opening angle. Figure \ref{fig:tj_tview} displays the mean $f_{b,{\rm eff}}$, averaged over $A_r$, density, and $z$ as a function of $\theta_{\rm jet}$ for the sGRB for different input sGRB fluence $S_{\rm bol}$ values. Faint sGRBs (bottom right of Figure \ref{fig:tj_tview}) tend to have $f_{b,{\rm eff}}<f_b$, because many of these sGRBs yield afterglows too faint to be detected. Oppositely, bright afterglows (top left of Figure \ref{fig:tj_tview}) can be detected beyond the edge of the top-hat jet -- and to increasingly large angles for Gaussian jets -- leading to larger effective beaming fractions. 

\begin{figure}[!h]
\centering
\includegraphics[width=1\columnwidth]{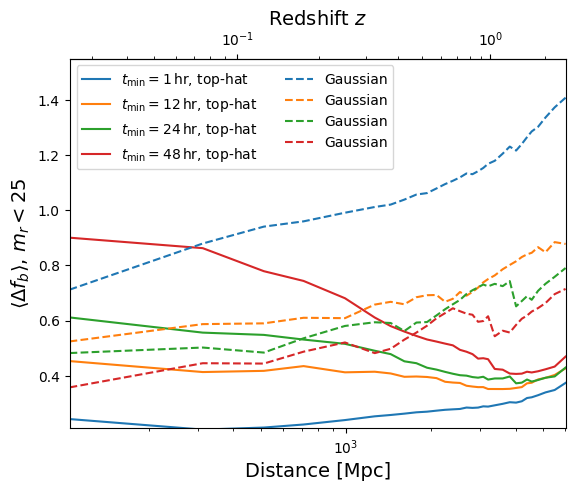}
\caption{The mean rate contribution from different jet regions, as a function of distance (or redshift) for afterglow observations reaching a depth of $m_r=25$ mag at different observations times.}
\label{fig:offaxis_rate_ratio}
\end{figure}

In Section \ref{sec:oa_rates} below, we determine rates for separate afterglow populations based on jet type and whether or not the jet is viewed within the top-hat core. To illustrate how this is done, Figure \ref{fig:offaxis_rate_ratio} shows the average effective beaming fraction as a function of distance for $m_r<25$ mag follow-up, factored into contributions $\Delta f_{b,{\rm eff}}$ corresponding to the fractional increases in rates that arise from the edge of the top-hat jet ($\Delta f_{b,{\rm top}}$) and the wings of a potential Gaussian jet ($\Delta f_{b,{\rm Gauss}}$), with:
\begin{equation}
f_{b,{\rm eff}} = (1+\Delta f_{b,{\rm top}}) (1+\Delta f_{b,{\rm Gauss}}) f_b.
\label{eq:offaxis_rate_ratio}
\end{equation}
The wings of the top-hat jet (solid curves in Figure \ref{fig:offaxis_rate_ratio}) contribute mostly for nearby events, because the afterglows are faint. The contribution increases with time, because afterglows rise at late time (see, e.g., the dotted curves in Figure \ref{fig:afterglow_param}). The Gaussian wings (dashed curves in Figure \ref{fig:offaxis_rate_ratio}) contribute more at higher redshift, because the afterglow is younger in rest-frame by $1+z$, and predominantly at early-time when the afterglow is sufficiently bright to permit viewing of the wings.

\begin{figure}[!h]
\centering
\includegraphics[width=1\columnwidth]{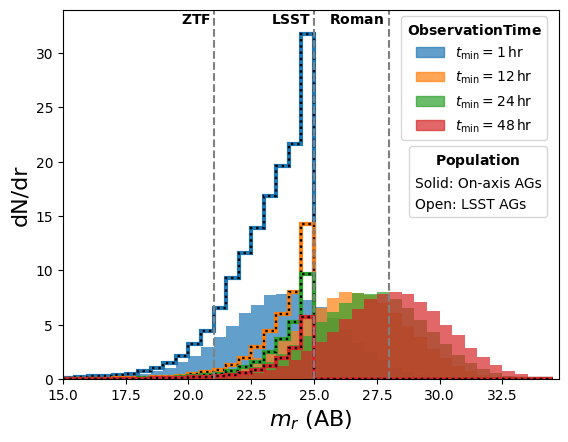}
\caption{The distribution of peak $r$-band apparent magnitudes at different observation times for the simulated sample of Swift/Fermi on-axis AGs (solid histograms), and the total LSST detectable ($m_{r} < 25$) afterglow population including both on-axis and orphan afterglows (open histograms). Optical detection limits in $r$-band are overlaid for ZTF, LSST, and Roman (gray, dashed).}
\label{fig:mag_dist}
\end{figure}

Figure \ref{fig:mag_dist} displays the predicted distributions in observed magnitude at different epochs, separated into on-axis only (solid histograms) and on- and off-axis components (open histograms) in the case of LSST. The solid histograms shift right in time as the afterglows decay in flux, with most of the population being fainter than $m_r=23$ after 1 hour and fainter than $m_r=28$ at 2 days after the sGRB. We note that these magnitudes are broadly consistent with those observed \citep[see, e.g.,][]{2011ApJ...734...96K}, considering that only the brightest $\sim$ 20\% of sGRBs afterglows -- corresponding to $m_r\lesssim 21.5$ mag at $1\, \rm hr$ in Figure \ref{fig:mag_dist} -- have historically been observed with sufficient brightness to determine redshifts \citep[see,][]{Dimitrova_2023}.

\section{\textbf{Afterglow Predictions}} \label{sec:result}
\subsection{Predicted Detectable Rates}\label{sec:oa_rates}
We compute the effective beaming fractions (Equation \ref{eq:fbmax}) for both jet models. We differentiate below between the on- and off-axis contribution by defining the on-axis contribution as that arising from Equation \ref{eq:fbmax} integrated from zero to $\theta_{\rm jet}$. The effective beaming fractions are then averaged over Galactic extinction and circumburst density.

We consider two classes of orphans: those of strongly collimated off-axis sGRBs, and those of sGRBs with prompt emission fainter than the $\gamma$-ray detector threshold. To obtain observable afterglow rates from the optical beaming fractions, we start from the number of $\gamma$-ray detectable sGRBs, $R_{\rm GRB}^{\rm det} = N_{\gamma}^{\rm det}/f_b$, which is larger than the number $N_{\gamma}^{\rm det}$ of detectable sGRBs from the world model pointed toward us above a Swift minimum signal-to-noise ratio corresponding roughly to a gamma-ray fluence threshold of $S_{\rm bol, min} = 10^{-8} \rm\, erg\, cm^{-2}$) (see \cite{Dimitrova_2023}). We also consider the events below the $\gamma$-ray detectability threshold, $R_{\rm GRB}^{\rm non-det} = (N_{\gamma}^{\rm tot}/f_b) - R_{\rm GRB}^{\rm det}$, where $N_{\gamma}^{\rm tot}$ is the total world-model sGRB sample. For each afterglow class, these sGRB rates are weighted by the corresponding $f_{b,\, \rm eff}$: 
\begin{equation}
R_{\rm AG} = {f_{b,\, \rm eff}}\, R_{\rm GRB}.
\label{eq:ag_propto}
\end{equation}

Off-axis orphan rates are defined for both jet models, while Faint orphans are derived from the total optically detectable Gaussian-jet population, thereby including both on- and off-axis events. The Faint orphan afterglows reflect the afterglows of all sGRBs below the Swift/Fermi detection threshold, regardless of orientation. 

The resulting rates are then summed over the physical parameters $\theta_i = (z, \theta_{\rm jet}, S_{\rm bol})$ as needed. In addition to including the Poisson error on the source counts, this procedure is carried out separately for 4000 draws from the MCMC world model -- thinned by a factor of 10 to guarantee statistical independence -- in order to calculate confidence intervals. In Table \ref{tab:OAtmin}, we report the predicted rates of optically detectable on-axis afterglows and a combined Off-axis and Faint orphan population, for a range of survey depths and observation times. In Table \ref{tab:OA_aLIGO}, we report predicted optically detectable afterglow rates for the upcoming aLIGO O5 run, showing the contributions from each afterglow class in optimistic LSST and Roman observing cases. 

\begin{table}[hbt!]
\renewcommand{\arraystretch}{1.3}
\centering
\caption{\textbf{Predicted Rate of Optically Detectable Afterglows (all-$\mathbf{z}$, all-sky, $\mathbf{yr^{-1}}$).} All uncertainties correspond to $90\%$ confidence intervals.}
\begin{tabular}{llll}
\hline
\hline
\multicolumn{4}{c}{\textbf{On-axis AGs (Swift/Fermi)}} \\
\hline
\hline
$\mathbf{t_{min}}$ & \textbf{ZTF} $m_r < 21$ & \textbf{LSST} $m_r < 25$ & \textbf{Roman} $m_r < 28$\\
\hline
$1\, \mathrm{h}$ & $7.5^{+2.8}_{-2.0}$ & $55^{+9}_{-8}$ & $83^{+12}_{-11}$\\
$12\, \mathrm{h}$ & $0.7^{+0.6}_{-0.3}$ & $18^{+5}_{-4}$ & $60^{+10}_{-9}$ \\
$24\, \mathrm{h}$ & $0.3^{+0.4}_{-0.2}$ & $11^{+4}_{-3}$ & $49^{+9}_{-8}$ \\
$48\, \mathrm{h}$ & $0.1^{+0.2}_{-0.1}$ & $6.0^{+2.5}_{-1.8}$ & $36^{+8}_{-6}$ \\
\hline
\hline
\multicolumn{4}{c}{\textbf{OAs (Off-axis \& Faint)}} \\
\hline
\hline
$\mathbf{t_{min}}$ & \textbf{ZTF} $m_r < 21$ & \textbf{LSST} $m_r < 25$ & \textbf{Roman} $m_r < 28$\\
\hline
$1\, \mathrm{h}$ & $9.5^{+4.8}_{-3.0}$ & $342^{+96}_{-79}$ & $2982^{+1128}_{-1119}$ \\
$12\, \mathrm{h}$ & $0.8^{+0.9}_{-0.4}$ & $34^{+15}_{-9}$ & $459^{+133}_{-110}$ \\
$24\, \mathrm{h}$ & $0.5^{+0.6}_{-0.3}$ & $18^{+9}_{-6}$ & $263^{+83}_{-56}$ \\
$48\, \mathrm{h}$ & $0.3^{+0.5}_{-0.2}$ & $8.9^{+5.9}_{-3.3}$ & $135^{+49}_{-29}$ \\
\hline
\hline
\end{tabular}
\label{tab:OAtmin}
\end{table}

Our highest predicted rate for detectable on-axis afterglows, $\approx 83^{+12}_{-11}\,\rm yr^{-1}$ (all-$z$, all-sky) from Roman at $t_{\rm min} = 1\, \rm hr$, is about $1.5\times$ higher than the comparable predicted rate within LSST's sensitivity limit. This factor increases at later times, as Roman-like sensitivity continues to detect fading events that have become too faint for current survey capabilities. In comparison to ZTF, the increased depths of LSST and Roman represent a substantial advancement in afterglow recovery. We estimate $\approx 2982^{+1128}_{-1119}$ Roman, and $342^{+96}_{-79}$ LSST intrinsically detectable OAs per year (all-$z$, all-sky) at $t_{\rm min} = 1\, \rm hr$. The ZTF model detects only a small fraction of these, approximately $3\%$ of the LSST yield (Table \ref{tab:OAtmin}). 

\begin{figure}[h!]
\centering
\includegraphics[width=1\columnwidth]{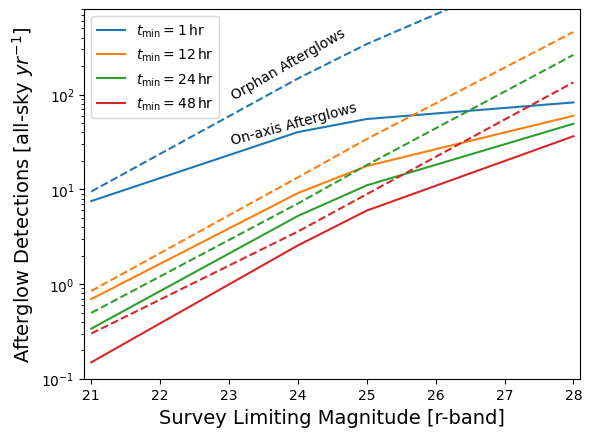}
\caption{The predicted all-sky per year afterglow detection rates as a function of $m_r^{\rm lim}$ for the considered follow-up observation times. Solid curves show the on-axis AGs, and dashed curves show the OAs - which include both the Off-axis and Faint orphan class.}
\label{fig:rates_maglim_oa}
\end{figure}

A steep decline in the predicted rates is observed with time, following the expected power-law decay typical of sGRB afterglows. After half a day, about two thirds of predicted on-axis afterglows are lost in LSST relative to those at $t_{\rm min} = 1 \rm hr$, while Roman retains $\approx 72\%$ of its population. Roman's greater depth provides a clear advantage at late times compared to other detectors - remaining capable of detecting a significant population of both on-axis and orphan afterglows after two days, by which point nearly all events have fallen below detection limits in ZTF, and only a few remain in LSST. 

\begin{figure*}[!t]
\centering
\includegraphics[width=\textwidth]{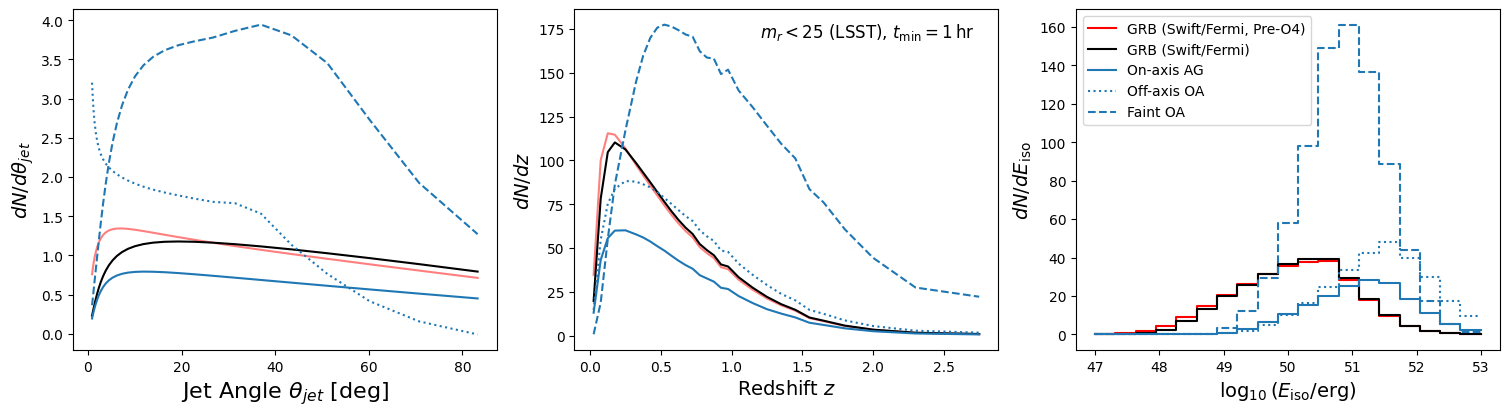}
\caption{The predicted rates of Swift/Fermi sGRBs prior to O4 analysis (red) and post-O4 result (black), on-axis afterglows (blue), Off-axis orphan afterglows (blue, dotted), and Faint orphan afterglows (blue, dashed) with respect to $\theta_{\rm jet}$ (left), $z$ (center), and $E_{\rm iso}$ (right). The sGRB rates are rescaled by a factor of $0.5$ for the purpose of visualization.}
\label{fig:obs_dist}
\end{figure*}

For every doubling $t_{\rm min}$ after $12\, \rm hr$, Roman and LSST detectable on-axis AG rates decrease by $\sim 1.2-1.8\times$, while orphan events decline more steeply, by roughly a factor of two. Figure \ref{fig:rates_maglim_oa} shows that increasing survey depth preferentially enhances the recovery of orphans, while later follow-up times shift peak detectability toward fainter, delayed afterglow emission. Thus, we emphasize that the capture of OAs requires fast follow-up alongside deep sensitivity. 

We note that at Roman's depth, rate prediction becomes uncertain, as such afterglows lie at the jet edge and faint end of the luminosity function below Swift/Fermi threshold - regimes that are weakly constrained. Their potential detection with Roman is therefore of significant interest for probing this poorly understood edge of parameter space, where event rates may be quite high. While the rates presented here reflect detectability based solely on magnitude criteria, in Section \ref{sec:obs} and Table \ref{tab:oa_det} we discuss and include detection rates from additional survey considerations.

\subsection{Predicted Observable Distributions}

Future OA detections will also help to constrain our world model and to inform our general understanding of sGRB properties. To illustrate, Figure \ref{fig:obs_dist} displays predicted distributions for $\theta_{\rm jet}$, $z$, and $E_{\rm iso}$, focusing on the optimistic
case of detections with the LSST at $t_{\rm min} = 1\,\rm hr$. Distributions are also shown for the pre- and post-O4 world model to illustrate the impact of aLIGO constraints.

The on-axis afterglows exhibit a relatively uniform distribution across $\theta_{\rm jet}$, with a mild enhancement around $\theta_{\rm jet} \approx 12^{\circ}$, suggesting that such afterglows may arise from a wide range of jet opening angles. In contrast, Off-axis orphans show a sharp preference for narrow jets, peaking at $\theta_{\rm jet} \approx 1^{\circ}$, as such highly collimated jets produce intrinsically brighter afterglows that are preferentially detected. The Faint orphan population allows broader beaming with a peak at $\theta_{\rm jet} \approx 37^{\circ}$ - though the overall orphan population remains biased to tracing strongly collimated events. Prior to O4, the Swift/Fermi sGRB and on-axis afterglows are similarly distributed over jetting. Incorporating O4 
shifts the inferred intrinsic jet distribution towards moderately broader jets, peaking near $\theta_{\rm jet} \approx 19^{\circ}$ (Figure \ref{fig:obs_dist}, left). 

We predict afterglow and sGRB distributions for redshift out to $z = 3$. This encompasses the usual range of sGRB detection and includes the expected peaks at $z \lesssim 0.2-0.5$. Beyond these peaks, the on-axis afterglow, Off-axis orphan, and sGRB populations decline rapidly with increasing redshift. Overall, the sGRBs are concentrated at lower $z$, with minimal change observed between the pre- and post-O4 analysis. On-axis afterglows peak similarly at $z \approx 0.3$, and while Off-axis orphans show a modest shift towards higher redshift. Faint orphans exhibit a broader redshift distribution with an extended high $z$ tail, reflecting their origin in the intrinsically missed sGRB population. This highlights that, in a sufficiently deep optical survey, Faint orphans tend to occur at greater redshifts as the detection volume increases (Figure \ref{fig:obs_dist}, center). 

In terms of the isotropic-equivalent energy release, the afterglow populations are concentrated at higher energies, peaking at $E_{\rm iso} \approx 10^{50.8}-10^{51.4}$ erg, relative to the sGRB population, which peaks at $E_{\rm iso} \approx 10^{50.5}$ erg. This emphasizes a strong detection selection effect which is especially present for Off-axis orphans, favoring more energetic events and therefore predominately sampling the highest-energy sGRBs. On-axis afterglows exhibit a more modest shift towards larger $E_{\rm iso}$ relative to observed sGRBs. The Faint orphans occupy the intermediate regime, tracing more moderate luminosities typical of sGRBs (Figure \ref{fig:obs_dist}, right). We note that the distributions corresponding to one day follow-up with Roman follow broadly similar trends, although with a modestly expanded detectable redshift and jet-angle range, as well as a reduced bias toward the most energetic events.

\subsection{Afterglow Dependence on Beaming and Other Parameters}\label{sec:param_dep} 
Within a conservative aLIGO O5 BNS range of $D< 225$ Mpc, based on current projections \citep{LVK}, we predict the rate of each afterglow class in favorable scenarios of LSST and Roman detection (Table \ref{tab:OA_aLIGO}). Off-axis orphans in a Gaussian jet model are most frequently detected, with a rate of $\approx 1.3-1.4$ all-sky per year in both detector configurations. In a top-hat model they are comparatively rare. This suggests that significant emission is present at larger viewing angles, enabling detection in sufficiently sensitive surveys. Consequently, higher observed Off-axis rates in the future with such surveys would support a structured jet model and thus constrain sGRB jetting. Although these observation cases do not generate a significant number of Faint orphans ($<1 \rm yr^{-1}$ in each scenario), their contribution may be quite large in future surveys with greater sensitivity.

\begin{table}[h!]
\centering
\renewcommand{\arraystretch}{1.5}
\caption{\textbf{Jet model Contribution to Predicted Optically Detectable Afterglow Rates (all-sky, $\bf yr^{-1}$) for $\bf D < 225$ Mpc (aLIGO O5)}. All uncertainties correspond to $90\%$ confidence intervals.}
\begin{tabular*}{\columnwidth}{@{\extracolsep{\fill}}lcc}
\hline
\hline
 \textbf{Afterglow}
& \textbf{LSST} $m_r < 25$ 
& \textbf{Roman} $m_r < 28$ \\
[-4pt]
 {\footnotesize Jet-Type}
& {\footnotesize $(t_{\rm min}=1 \rm hr)$}
& {\footnotesize $(t_{\rm min}=24 \rm hr)$} \\
\hline
\hline
On-axis AG & $1.1\pm0.4$ & $1.0^{+0.4}_{-0.3}$ \\
Off-axis OA, Top-hat & $0.2\pm 0.1$ & $0.6^{+0.3}_{-0.2}$ \\
Off-axis OA, Gaussian & $1.3^{+0.5}_{-0.4}$ & $1.4^{+0.5}_{-0.4}$ \\
Faint OA & $0.3\pm0.2$ & $0.2\pm0.1$ \\
\hline
\hline
\end{tabular*}
\label{tab:OA_aLIGO}
\end{table}

We note that the multi-messenger implication of afterglow detection is dependent on assumed jet structure. Within the GW significant volume considered, a Gaussian jet model favors Off-axis orphans as potential EM counterparts to aLIGO events, while a top-hat model predicts that on-axis afterglows are the most likely candidates. In practice, additional observational limitations must be considered. These are discussed in detail in Section \ref{sec:aLIGO_implication}.

\begin{figure}[h!]
\centering
\includegraphics[width=1\columnwidth]{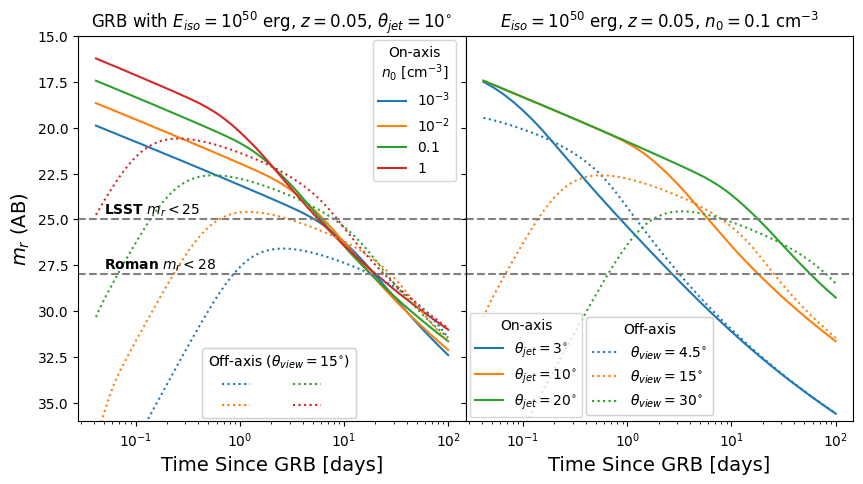}
\caption{Simulated sGRB $r$-band afterglow light curves in a top-hat jet model, assuming fixed values corresponding to a favorable detection scenario, for on-axis (solid) and off-axis (dotted) viewing angles. LSST and Roman $r$-band detection thresholds are overlaid (gray, dashed). \textit{Left:} $E_{\rm iso}$, $z$, and $\theta_{\rm jet}$ are fixed to show light curve dependence on density. \textit{Right:}  $E_{\rm iso}$, $z$, and $n_0$ are fixed to show light curve dependence on jet geometry.}
\label{fig:afterglow_break}
\end{figure}

In Figure \ref{fig:afterglow_break}, we display afterglow light curves generated with \texttt{AfterglowPy} for sGRBs with fixed parameters chosen to favor detectability, for on- and off-axis viewing cases. Afterglows with narrow jets ($\theta_{\rm jet} = 3^{\circ}$ here) are brighter and peak earlier, requiring fast follow-up of within a few days. Wider jets extend the detectability window for on-axis events. For off-axis viewing, increasing $\theta_{\rm view}$ shifts the peak emission to later times, but with a significant decrease in brightness. Observed afterglow populations are therefore expected to be biased towards on-axis events, as recovering narrow jets requires early follow-up and detecting off-axis events requires extremely deep sensitivity. 

sGRBs viewed off-axis are especially sensitive to density, as a larger $n_0$ corresponds to brighter peak emission that is detectable for longer. In varying the density of the considered sGRB, most models fall below LSST sensitivity by $\gtrapprox 10$ days, and below Roman limits around $\gtrapprox 20$ days. As environmental properties strongly bias the detectable afterglow population, future detections are likely to occur in denser merger environments.

The LSST can primarily detect on-axis afterglows, especially in high-density environments. Roman depth extends this sensitivity to off-axis events and lower densities, across a wider range of viewing angles, and allows off-axis detection at earlier times. For example, the optical emission of the considered off-axis sGRB with $n_0 = 10^{-3}\,\rm cm^{-3}$ (Figure \ref{fig:afterglow_break}, left) does not exceed the LSST threshold, but is seen in Roman for several weeks. Similarly, a survey reaching Roman depth could observe the shown off-axis afterglow with $\theta_{\rm jet} = 20^{\circ}$ (Figure \ref{fig:afterglow_break}, right) before a day post-burst up to several months, whereas the same event becomes detectable in LSST after approximately two days and remains observable for only a few days. While prompt response is advantageous in capturing the early emission of on-axis afterglows and the rising phase of off-axis events, deeper surveys may recover late-time jet-breaks and decay behavior over extended timescales.

\section{\textbf{Afterglow Observation Strategy and Feasibility}} \label{sec:obs}

\subsection{Predicted Survey Detection Rates}  
In practice, afterglow detection is additionally dependent on the sky coverage and cadence of the observing survey. We estimate the instantaneous annual detection rate for a given survey as a function of observation time:
\begin{equation}\label{eq:cad}
    R_{\rm det}(t_{\rm min}) = R_{\rm AG}(t_\mathrm{min}, m_r^{\rm lim})\,f_{sky}\,(\frac{t_{\rm min}}{t_{\rm cad}})
\end{equation}
Where $f_{sky}$ is the fraction of the sky imaged by the survey, and $t_{\mathrm{cad}}$ is survey cadence in hours. 
We then compute time averaged detection rates over the cadence window $t_0 \le t_{\rm min} \le t_{\rm {cad}}$ (for which we adopt $t_0 = 1\, \mathrm{hr}$) for the LSST Wide Fast Deep (WFD) survey \citep{Ivezic2019} with:
\begin{equation}\label{eq:int_detect}
R_{\mathrm{det}}^{\rm cad} = \frac{f_{\mathrm{sky}}}{t_{\rm cad}}\, \int_{t_0}^{t_{\rm cad}} R_{\rm AG}(t_\mathrm{min}, m_r^{\rm lim}) \, dt_{\rm min} 
\end{equation}
We assume an approximately half-sky WFD footprint, adopting the upper end of estimates ($20,000\, \rm deg^2$), and a representative cadence of $t_{\rm cad} = 72\, \rm hr$ consistent with the expected median revisit times \citep{Ivezic2019, Jones2020} to predict a total detection rate of $\approx 5.3^{+1.7}_{-1.2}$ on-axis and $\approx 11^{+5}_{-3}$ orphan afterglows in WFD. 
We find peak detection at $\sim t_{\mathrm{min}} =  48\,\mathrm{hr}$ for on-axis afterglows in WFD, indicating sensitivity primarily to post-peak evolution and slow decay. Off-axis afterglows dominate the detected WFD population and increase with $t_{\rm min}$, suggesting preference for their slower evolving, late peaking emission, whereas Faint orphans fade rapidly below optical limit beyond $t_{\rm min} = 1\, \rm hr$.

With a simple afterglow decay model $R^{\rm cad}_{\rm det} \propto t^{-1}$, Equation \ref{eq:int_detect} gives the approximate scaling
$R_{\rm det}^{\rm cad} \propto \log(t_c)/t_c$ for some new cadence $t_c$. We apply this with survey specific parameters to estimate the afterglow detection rates for Roman's High Latitude Time Domain Survey (HLTDS) \citep{Hounsell2023}, and an idealized survey. Despite Roman's increased sensitivity, the predicted total afterglow detection rate of HLTDS remains low, primarily due to it's limited sky coverage - yielding only $\approx 0.1$ on-axis and orphan afterglows combined per year, assuming the expected baselines for a HLTDS footprint of $\sim 18$ deg$^2$ and cadence $t_{\mathrm{c}} = 120\, \mathrm{hr}$ \citep{RomanHLTDS, HLTDSCadence}. In HLTDS, the rates for both on-axis afterglows and Off-axis orphans rise over $t_{\rm min}$, where Off-axis orphans account for most detections at later times, while Faint orphans dominate across $t_{\rm min}$. The predicted cumulative real-time survey detection rates are presented in Table \ref{tab:oa_det}.

\begin{table}[hbt!]
\renewcommand{\arraystretch}{1.2}
\centering
\caption{\textbf{Predicted Optical Survey Cumulative Afterglow Detection Rates ($\bf yr^{-1}$)}} 
\begin{tabular}{ll}
\hline
\hline
\multicolumn{2}{c}{\textbf{LSST WFD}} \\
\multicolumn{2}{c}{$\bf m_r < 25\,$ $\bf f_{\rm \bf sky} \approx 0.5\,$  $\bf t_{\rm \bf cad} = 3\, \rm \bf d$} \\ 
\hline \noalign{\vskip 4pt}
On-axis: $5.3^{+1.7}_{-1.2}$ & Orphan: $11^{+5}_{-3}$ \\ \noalign{\vskip 4pt}
\hline
\hline
\multicolumn{2}{c}{\textbf{Roman HLTDS}} \\
\multicolumn{2}{c}{$\bf m_r < 28\,$ $\bf f_{\rm \bf sky} \approx 4\times10^{-4}\,$ $\bf t_{\rm \bf cad} = 5\, \rm \bf d$} \\
\hline \noalign{\vskip 4pt}
On-axis: $0.013\pm 0.002$ & Orphan: $0.087^{+0.025}_{-0.020}$ \\ \noalign{\vskip 4pt}
\hline
\hline
\multicolumn{2}{c}{\textbf{Ideal Survey}} \\
\multicolumn{2}{c}{$\bf m_r < 28\,$ $\bf f_{\rm \bf sky} \approx 0.5\,$ $\bf t_{\rm \bf cad} = 1\, \rm \bf d$} \\
\hline \noalign{\vskip 4pt}
On-axis: $51^{+9}_{-8}$ & Orphan: $331^{+95}_{-78}$ \\ \noalign{\vskip 4pt}
\hline
\hline
\end{tabular}
\label{tab:oa_det}
\end{table}

We note that other factors may impact afterglow detectability. First, WFD employs multi-filter sampling, allocating $\approx 20\%$ of visits to each of the primary transient relevant \textit{r}-,\textit{i}-, and \textit{z}-band filters \citep{Jones2020, Ivezic2019}. This limitation is less severe for HLTDS, where all filters contribute to afterglow detectability \citep{Roman}. 

Furthermore, there is uncertainty in discerning orphans from other transients, such as fast evolving kilonovae (KN), tidal disruption events, fast blue optical transients, and Type Ic or Type II-P/L supernovae (SNe) \citep{Metzger2017, Gezari2021, Ho2023FBOT, Filippenko1997, Anderson2014}. Rate estimates indicate that afterglows will be comparatively rare to these other transients in WFD and HLTDS \citep{sn_rate, Pierel2020RomanSNe, Perkins2024, Karmen2026TDE, Ho2023FBOT}. Additionally, LSST is strongly dependent on ground-based observing conditions, and science cases are limited by scheduling restrictions. 

\subsection{Roman, LSST Relative Performance and Prospects}\label{subsec:survey_comp}
\begin{figure*}[!t]
\centering
\includegraphics[width=\textwidth]{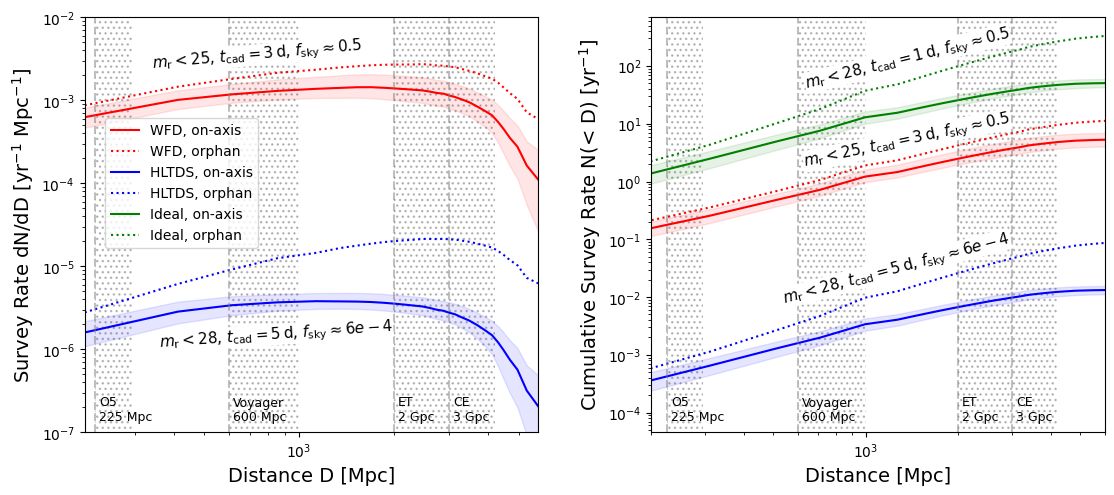}
\caption{\textit{Left}: Predicted rate of on-axis (solid) and orphan (dotted) afterglow detection for WFD (red) and HLTDS (blue) as a function of distance. \textit{Right}: Predicted cumulative rate of on-axis (solid) and orphan (dotted) afterglow detection for an ideal detector (green), WFD (red), and HLTDS (blue). We overlay the representative BNS sensitivity ranges for aLIGO O5, aLIGO Voyager \citep{Voyager2023, LIGO_Voyager}, the ET \citep{et_volume}, and the CE \citep{Hall2022CE}. $90\%$ confidence intervals for the on-axis rates are shown by the shaded regions.}
\label{fig:afterglow_dndd}
\end{figure*}

In Figure \ref{fig:afterglow_dndd}, we plot the predicted on-axis and orphan afterglow yields of WFD, HLTDS, and an ideal detector - where WFD proves to be vastly more capable of afterglow detection than HLTDS. Although the higher cadence of WFD increases the rate comparatively to HLTDS, this benefit is limited by survey depth, as Faint orphans and early-time emission often remain below detection thresholds despite more frequent imaging. Conversely, a slower cadence survey may miss early detection, but can still capture extended afterglow decay and the delayed peaks of Off-axis orphans given HLTDS sensitivity. The relative fraction of afterglow type in a detected population is therefore more strongly governed by survey sensitivity. This effect becomes prominent for larger detection volumes, as the HLTDS model shows increasing bias for orphans with greater distance, while the population fractions in WFD remain more consistent (Figure \ref{fig:afterglow_dndd}). Afterglow detection rates scale linearly with survey sky-area, giving WFD a significant advantage due to it's wide field coverage.

An example ideal detector is modeled on the sensitivity of Roman, the survey sky-coverage of WFD, and a reasonably fast cadence of $t_{\mathrm{cad}} = 1$ day. Going from a WFD cadence of 3-days to a 1-day cadence improves the cumulative afterglow detection rates by a factor of $\approx 2$, while extending the depth of a WFD like survey to that of HLTDS yields an $\approx 4 \times$ increase in on-axis and $\approx 13 \times$ increase in orphan afterglow detectable rates. The ideal experiment thus considerably raises detection efficiency, predicting $\approx 51^{+9}_{-8}$ on-axis, and $331^{+95}_{-78}$ orphan afterglows per year. 
 
Although HLTDS is expected to operate entirely within the WFD timeline, no significant benefit in joint real-time detection rates is expected due to the low probability of afterglow detection in HLTDS and the lack of coordinated observing strategy or overlapping sky coverage \citep{Ivezic2019, Hounsell2023, RomanHLTDS}. The telescopes may reasonably collaborate in triggered observation, with the most realistic pathway being LSST detection of a transient prompting a ToO request for deeper imaging with Roman. Targeted follow-up of Swift alerts, \citep{Gehrels2004} may be easily contained in LSST and Roman's FOV's \citep{Ivezic2019, Roman}. The combination of their archival data could anticipate benefit as a multi-survey time-domain strategy, with LSST being advantageous for on-axis events and variability history; and Roman for faint-source detection and deeper follow-up of afterglow environment.

\section{\bf Implication for LIGO and Future Experiments}\label{sec:aLIGO_implication}
\subsection{Afterglow Multi-messenger Potential}
HLTDS and WFD should considerably overlap with the planned timeline of aLIGO O5 \citep{Roman, Ivezic2019, LVK}, with Table \ref{tab:OA_aLIGO} showing that afterglows are a possible channel for joint optical-BNS detection and for probing the GW-relevant BNS population. Accounting for all the considered optical survey limitations (Section \ref{sec:obs}), and assuming a shared 2.5 year observing duration, we find a $66\%$ probability of at least one real-time WFD/HLTDS afterglow detection during O5 within the adopted BNS range. WFD may have several years of overlap with the aLIGO Voyager upgrade \citep{Voyager2023} - yielding a more promising $\approx 94\%$ probability of at least one WFD associated afterglow observation during this era, adopting 2 years of shared observing time and assuming the lower end of the adopted Voyager BNS range ($\approx 600\, \mathrm{Mpc}$). 

The timelines of WFD and HLTDS are expected to have negligible intersection with proposed third-generation GW detectors, the Einstein Telescope (ET) and the Cosmic Explorer (CE) \citep{Maggiore2020ET, Hall2022CE}. Considering an optimistic scenario of a future optical surveyor in this era that mirrors our ideal detector, we predict $\approx 125$ and $216\, \mathrm{yr}^{-1}$ afterglow detections within the adopted lower BNS ranges of ET ($\sim 2\, \mathrm{Gpc}$) and CE ($\sim 3\, \mathrm{Gpc}$). 

\subsection{sGRB-BNS Rate}
As seen in Figure \ref{fig:grb_rate}, the post-O4 observed rate of Swift/Fermi sGRBs closely tracks the underlying sGRB population out to a distance comparable to the upper O4 BNS range. Beyond this, the true all-sky sGRB rate increasingly exceeds this detected population due to selection effects. Including O4 places tighter constraints on the observed sGRB population model, leading to increased separation from the inferred BNS rate. Our model is consistent with the two observed BNS events from O2+O3, within their uncertainty intervals. Table \ref{tab:grb_rate} displays our cumulative sGRB and BNS rates as a function of distance. 

\begin{figure}[hbt!]
\centering
\includegraphics[width=1\columnwidth]{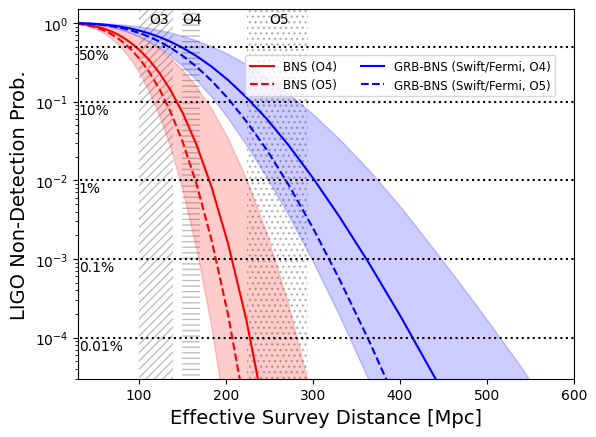}
\caption{The probability of non-detection of a BNS merger in aLIGO O4 (red) and O5 (red, dashed), and a BNS merger with an associated Swift/Fermi sGRB in O4 (blue) and O5 (blue, dashed). The representative BNS ranges of aLIGO O3, O4, and O5, are overlaid (shaded regions).}
\label{fig:grb_bns_prob}
\end{figure}

In Figure (\ref{fig:grb_bns_prob}) we evaluate the probability of our rates producing the non-detection result of aLIGO O4, and again during O5. The achieved BNS range of O4 is now thought to be $\approx 150 - 170$ \citep{LVK}, over which our model predicts a $\approx 3-7\%$ probability of O4 BNS non-detection, with credible intervals spanning $\approx 0.1 - 25\%$. Our results could suggest that the BNS sensitivity range of O4 may be lower than expected. Alternatively, the true BNS rate may lie on the conservative end of current estimates ($\sim 10-1000\, \rm Gpc^{-3}\, yr^{-1}$), as independently constrained by sGRB populations \citep{Coward2012}, BNS population-synthesis models \citep{Dominik2012}, and gravitational-wave population analyses \citep{Abbott_2025_GWTC4_Population}. 

The probability of aLIGO O5 not detecting a BNS merger is near zero in our model. For BNS mergers with an associated sGRB, the corresponding non-detection probability is $\approx 0.25 - 5\%$ across the adopted O5 BNS range, assuming an effective live time of $2.5$ years for O5 \citep{LVK}. Given accurate aLIGO BNS sensitivity, continued non-detection would strongly suggest a lower intrinsic BNS merger rate, potentially requiring increased contribution from an alternative sGRB progenitor channel.

\begin{table}[hbt!]
\renewcommand{\arraystretch}{1.3}
\centering
\caption{\textbf{Predicted Cumulative sGRB and BNS Rates (all-sky, $\rm \bf yr^{-1}$)} in lower end BNS detection ranges. Error regions correspond to $90\%$ confidence intervals.}
\begin{tabular}{llll}
\hline 
\hline
\textbf{D}  & \textbf{sGRB} & \textbf{Incl. faint} & \textbf{BNS} \\
\hline
\hline
3 Gpc (CE) & $68^{+12}_{-11}$ & $3520^{+3160}_{-1730}$ & $12400^{+10500}_{-5800}$ \\
2 Gpc (ET) & $47^{+11}_{-10}$ & $1040^{+880}_{-500}$ & $3650^{+3040}_{-1710}$ \\
600 Mpc (Voyager) & $9.7^{+3.8}_{-3.1}$ & $26^{+22}_{-12}$ & $91^{+77}_{-43}$ \\
225 Mpc (O5) & $1.2^{+0.7}_{-0.5}$ & $1.3^{+1.1}_{-0.6}$ & $4.7^{+3.9}_{-2.2}$ \\
150 Mpc (O4) & $0.4^{+0.3}_{-0.2}$ & $0.4^{+0.3}_{-0.2}$ & $1.4^{+1.2}_{-0.7}$ \\
100 Mpc (O2-O3) & $0.1\pm 0.1$ & $0.1 \pm 0.1$ & $0.4^{+0.3}_{-0.2}$ \\
\hline
\hline
\end{tabular}
\label{tab:grb_rate}
\end{table}

\newpage
\section{\textbf{Conclusions}}\label{sec:conclusions}
We have predicted probabilistic rates and observable distributions of sGRBs and their afterglows for the LSST and Roman optical surveys, from a sample of sGRBs generated from source BNS mergers (Figure \ref{fig:grb_rate}). 
For opportunistic cases of observation, at $t_{\rm min} = 1\rm\, hr$ with LSST, we predict $55^{+9}_{-8}$ on-axis afterglows and $342^{+96}_{-79}$ OAs, and for Roman at $t_{\rm min} = 24 \rm\, hr$, $49^{+9}_{-8}$ on-axis afterglows and $263^{+83}_{-56}$ OAs (all-$z$, all-sky, $\rm yr^{-1}$), within optical threshold (Table \ref{tab:OAtmin}). 

Off-axis orphans in the LSST model preferentially sample the highest energy sGRBs, with strong selection towards narrow, energetic jets demonstrating observational bias. Those of on-axis events are detectable across a wider jet range, and probe a relatively lower end of the luminosity function, providing a comparatively more representative sample of the underlying sGRB population. The LSST Faint orphans allow moderate energetics with wider jet angles, showing the broadest $z$ reach, which indicates tracing of intrinsically missed sGRBs. The modeled Roman afterglow population displays similar trends, with less bias towards energetic events and somewhat extended redshift and jetting ranges (\ref{fig:obs_dist}). Our Swift/Fermi selected sGRB afterglows are broadly consistent with the Swift-era afterglow inferred constraints from \cite{2011ApJ...734...96K} and \cite{Fong2015}, reproducing characteristic bias for moderate jet opening angles, low redshift, and extension into the low energy regime. 

Afterglow light curve simulation with \texttt{AfterglowPy} demonstrates how optical afterglow detectability is governed jointly by circumburst density and jet geometry (Figure \ref{fig:afterglow_param}, \ref{fig:afterglow_break}). Consequently, survey sensitivity strongly shapes the observed population and inferred constraints. In LSST threshold, we expect observational bias for afterglows from denser environments - which produce brighter events, and from on-axis orientations - as the the narrow jet preference of off-axis sGRBs requires early-up follow-up or deeper survey sensitivity. The expected population above a Roman optical limit is more diverse, allowing for more off-axis events, wider $\theta_{\rm view}$, and lower circumburst density. Thus, Roman-like depth increases the accessible parameter space for detecting orphans, over extended timescales. 

We consider the additional limitations of the LSST and Roman transient surveys (Section \ref{sec:obs}) and estimate $5.3^{+1.7}_{-1.2}$ on-axis and $11^{+5}_{-3}$ OA detections per year in the LSST WFD survey, far outnumbering the detections in Roman's HLTDS (Table \ref{tab:oa_det}). Figure \ref{fig:rates_maglim_oa} measures how the rate of afterglow detection grows with survey depth - showing that Roman-like sensitivity benefits orphans more than on-axis events, while Figure \ref{fig:afterglow_dndd} shows the detection rates under various survey configurations - demonstrating the advantage of WFD's higher cadence and greater sky-coverage. However, the deep sensitivity of Roman is well suited for afterglow observation in follow-up of Swift triggers, which appears more feasible than real-time searches. 
We find that a deeper ($m_r^{\rm lim} \approx 28$), wider ($f_{\rm sky} \approx 0.5$) survey with a daily cadence, which we model in Section \ref{subsec:survey_comp}, is a realistically optimal design for real-time detection to obtain a complete afterglow sample and recover full evolutionary phases. Such an ideal survey could capture on-axis afterglows from over half of all Swift/Fermi sGRBs within $1\, \rm Gpc^{3}\, yr^{-1}$, while LSST and Roman combined could expect retrieval from roughly $6\%$ in this volume. 
    
The predicted rates of LSST and Roman discoverable afterglows within the adopted O5 BNS range are low ($R_{\rm AG}< 1.4$ for each jet model and observing scenario, Table \ref{tab:OA_aLIGO}), indicating that they are a potential, though unlikely EM counterpart to BNS-GW events in O5. Although aLIGO's Voyager update expects improved BNS sensitivity \citep{Voyager2023}, optical survey timing and observational constraints prove challenging (see Sections \ref{sec:obs}, \ref{sec:aLIGO_implication}). For the most optimistic scenario that adopts the upper end of projected aLIGO Voyager BNS ranges ($\sim 1000 \rm\, Mpc$ \citep{LIGO_Voyager}) and a two year overlap with WFD, we predict $\sim 3$ WFD afterglow detections during the Voyager era. 
A realistic framework to maximize optical counterpart detection could include a survey with Roman depth, fast response (about minutes to hours) and a single-pointing coverage suitable for GW-BNS follow-up ($\sim 10\, \mathrm{deg}^{2}$). This would increase the likelihood of joint-detection with third-generation GW-detectors, which promise better localizations ($\sim 0.1-1\, \mathrm{deg}^{2}$) \citep{Chan2018Localization} and greater sensitivity \citep{et_volume, Evans2021HorizonStudy, Hall2022CE}. 

Kilonovae may be a more promising channel to optical observation in upcoming LIGO runs (with predicted rates of about a few to a few tens, \citep{shah, Perkins2024}). We note that our predicted afterglow rates are lower than those of \cite{Perkins2024}, which may reflect differences in the assumed source population and luminosity function, potentially favoring intrinsically bright detectable events.

LSST may yield an afterglow sample large enough to provide an independent rate estimate of the underlying population and place meaningful constraints on the beaming correction - which is crucial for interpreting the tension between sGRB and GW-BNS merger rates. As jet structure strongly affects the predicted orphan rates, with a Gaussian jet model producing significantly more Off-axis orphans than a top-hat model (Table \ref{tab:OA_aLIGO}); such future orphan detections could directly constrain sGRB jet structure. Future GW observations may refine these constraints, as the absence of a confident BNS detection during O4 favors broader inferred sGRB jet opening angles, allowing the observed sGRB rate to be reproduced with fewer underlying BNS mergers.

An observed afterglow sample could allow distinction from other transients through the mapping of distributions (such as in Figure \ref{fig:obs_dist}), and could probe sGRBs in GW-BNS significant volumes. Completely ruling out the afterglows of long-GRBs in an observed population is unlikely, though solid scientific merit also exists in studying such orphans. As OAs trace the missed sGRB population, a WFD recovered sample may additionally restrict other fundamental sGRB properties. 

\bibliographystyle{aasjournal}
\bibliography{bib}

\end{document}